\documentstyle{elsart}
\def\gtrsim{ \> \lower4pt\hbox{${\buildrel > \over \sim}$} \> }
\def\lesssim{ \> \lower4pt\hbox{${\buildrel < \over \sim}$} \> }

\begin{document}
\begin{frontmatter}

\title{Prompt and Delayed High-Energy Emission \\
from Cosmological Gamma-Ray Bursts}

\author[Rice,NRL]{Markus B\"ottcher}
\author[NRL]{\& Charles D. Dermer}

\address[Rice]{Space Physics and Astronomy Department, 
Rice University, MS 108 \\
6100 S. Main Street, Houston, TX 77005-1892}
\address[NRL]{E. O. Hulburt Center for Space Research, Code 7653 \\
Naval Research Laboratory, Washington, DC 20375-5352}

\begin{abstract}

In the cosmological blast-wave model for gamma ray bursts (GRBs),
high energy ($\gtrsim 10$~GeV) $\gamma$ rays are produced either
through Compton scattering of soft photons by ultrarelativistic
electrons, or as a consequence of the acceleration of protons to
ultrahigh energies. We describe the spectral and temporal characteristics 
of high energy $\gamma$ rays produced by both mechanisms, and 
discuss how these processes can be distinguished through observations with 
low-threshold \v Cerenkov telescopes or GLAST. We propose that
Compton scattering of starlight photons by blast wave electrons can
produce delayed flares of GeV -- TeV radiation.

\end{abstract}
\end{frontmatter}

\section{Introduction}

The blast-wave model has met with considerable success in explaining 
the X-ray, optical and radio afterglows of GRBs \cite{mes97}. The evolving
blast wave accelerates particles to  ultrarelativistic energies and has
been suggested as the  source of ultrahigh-energy cosmic rays (UHECRs)
\cite{wax95,vie95}.  GeV -- TeV photon production, due primarily to
proton synchrotron radiation from UHECR acceleration in
GRBs during the afterglow phases, has recently been
predicted \cite{vie97,boe98}. A large Compton luminosity when
relativistic electrons scatter soft photons is expected as well. 
One source of seed photons is the electron synchrotron radiation, 
and the synchrotron self Compton (SSC) process has been calculated 
by \cite{chi98}. External soft photon sources could also be important 
in the Compton scattering process. Models for the GRB origin such as 
the hypernova/collapsar \cite{pac98,woo93} scenario suggest that the
sources of GRBs are associated with star-forming  regions and might 
thus be embedded in a starlight radiation field. Here we examine the
Compton scattering of starlight by relativistic electrons in the blast
wave.

\section{Compton Emission}

We compare high energy $\gamma$-ray emissions due to SSC and SSL 
(Scattered StarLight) processes, which can be estimated by comparing 
the comoving blast-wave frame energy densities $u'_{SL}$ and $u'_{sy}$ 
of starlight and synchrotron photons, respectively.  
Knowledge of the peak energy of the seed photon spectrum, 
which determines whether Compton scattering is suppressed due to
Klein-Nishina effects, is important for this comparison. The relativistic 
electrons are assumed to be distributed according to a power-law with
index $\gtrsim 3$ and a low-energy cutoff $\gamma_{\rm e, min} = 
\xi_e \, ( m_p / m_e) \, \Gamma$ \cite{mes94}, where 
$\xi_e = 0.1 \, \xi_{e, -1}$ is an electron equipartition 
factor, and $\Gamma = 300 \, \Gamma_{300}$ is the bulk 
Lorentz factor of the blast wave. The magnetic field is 
$H = \sqrt{8 \pi \, r \, m_p c^2 \, n_0 \, \xi_H} 
\Gamma$, where $r = 10 \, r_1$ is the shock compression 
ratio, $n_0$ is the number density of external
matter, assumed uniform, and $\xi_H = 10^{-6} \, \xi_{H, -6}$ is a
magnetic-field  equipartition factor. A value of $\xi_H \sim 10^{-6}$ is
required  to produce synchrotron spectra resembling observed
spectra in the prompt phase of GRBs \cite{chi98}.

\subsection{Synchrotron and SSC Processes}

The peak photon energy of the synchrotron spectrum
is $\epsilon'_{sy} \simeq  10^{-4} \Gamma_{300}^3 \, n_0^{1/2} 
\, q_{-4}$, where $q = \sqrt{r_1 \, \xi_H} \, \xi_e^2 = 10^{-4} 
\, q_{-4}$ is a combined equipartition factor. The energy 
density of the synchrotron radiation field may be estimated
using $u'_{sy} \approx \tau_T \, \gamma_{\rm e, min}^2 \,
u'_{H}$, where $\tau_T$ is the radial Thomson depth of the
blast wave. Specifying the width of the shell, $\Delta x'
\approx x' / \Gamma$, at the deceleration radius, we find
\begin{equation}
u'_{sy}\;
[{\rm erg 
\> cm}^{-3}] \simeq 7.3 \cdot 10^{-6} \, r_1^2\, n_0^{5/3} \xi_{H, -6}
\, E_{52}^{1/3} \, \xi_{e, -1}^2 \; \Gamma_{300}^{10/3}\; ,
\end{equation}
where $E_0 = 10^{52} \, E_{52}$~erg is the kinetic energy of the 
blast wave. SSC scattering can occur efficiently in the Thomson 
regime if $\xi_{e, -1} \, \Gamma_{300}^4 \, n_0^{1/2} \, q_{-4} \lesssim
1$. In this case, the SSC spectrum in the observer's frame peaks at
\begin{equation}
\epsilon_{\rm SSC} = h\nu_{\rm SSC}/m_ec^2 \simeq 10^8 \, \Gamma_{300}^6 \,
\xi_{e, -1}^2
\, n_0^{1/2} \, q_{-4} \, / \, (1 + z),
\end{equation}
which cannot exceed $\epsilon_{\rm IC, max} \simeq 1.6 \cdot 10^7 \, 
\xi_{e, -1} \, \Gamma_{300}^2/(1+z)$.

\subsection{Scattered Starlight}

We assume that the blast wave approaches a star of luminosity 
$L_{\ast} = \l \, L_{\odot}$ at a distance $d = d_{15}$~cm from a 
given point within the blast wave. Then the energy density of the
stellar radiation field in the blast-wave frame is
\begin{equation}
u'_{SL} [{\rm erg \> cm}^{-3}] \simeq {1.3\cdot 10^{-3} \, 
\l\Gamma_{300}^2 \over \, d_{15}^{2} }
\end{equation}
\cite{ds94}. By comparing eqs.\ (1) and (3), we see that
$u'_{SL}$ dominates $u'_{sy}$ when $d_{15} \lesssim d_{SSL}
\equiv 13 \, \l^{1/2} \, \Gamma_{300}^{-2/3} \, r_1^{-1} \, 
n_0^{-5/6} \, \xi_{H, -6}^{-1/2} \, \xi_{e, -1}^{-1} \, 
E_{52}^{-1/6}$.

If the starlight spectrum is
approximated by a thermal  blackbody of temperature
$T_0$[eV], then this spectrum peaks in the comoving blast-wave frame at
dimensionless photon energy $\epsilon'_{SL} \cong 1.6 \cdot 10^{-3} \,
\Gamma_{300} \, T_0$. Compton scattering can occur in the Thomson regime if
$\Gamma_{300}^2 \, \xi_{e, -1} \, T_0 \lesssim 10^{-2}$,
indicating that the SSL process becomes efficient 
only if $\Gamma \lesssim 30 \, (\xi_{e, -1} \, T_0)^{-1/2}$. 
This can be realized in dirty fireballs \cite{dcb99} or in the later
afterglow phase of blast-wave deceleration. When the 
above condition is satisfied, the SSL spectrum in the observer's frame
peaks at
\begin{equation}
\epsilon_{SSL} \cong 1.5 \cdot 10^5 \, \Gamma_{30}^4 
\, T_0 \, \xi_{e, -1}^2 \, / \, (1 + z),
\end{equation}
where $\Gamma_{30} = \Gamma/30$. The SSL
component can dominate once scattering occurs primarily in the Thomson
regime, at which point it is  likely to be the dominant electron
radiation mechanism in  the 10 -- 100~GeV regime. 

The minimum duration of flares from the SSL process is $\Delta 
t_{\rm min}\approx d/(\Gamma^2 c) \approx 40 d_{15}/\Gamma_{30}^2$~s.
The fraction of energy in GeV -- TeV photons produced by the SSL 
mechanism rather than through other processes is roughly given by
$[d_{SSL}/(x_{bw}/\Gamma)]^2$, which represents the fraction of the 
observable blast wave area where $u'_{SL}$ dominates $u'_{sy}$. 
The term $x_{bw}$ is the distance of the blast wave from the
explosion site. \v Cerenkov  telescopes with threshold energies 
below $\sim$~100~GeV and the planned GLAST satellite might be 
able to detect the predicted SSL radiation from nearby GRBs if 
they are indeed associated with star-forming regions. 

\section{Comparison with Gamma Rays from Hadronic Processes}

The high energy $\gamma$-ray signatures of UHECR acceleration in GRB 
blast waves have been investigated in detail in \cite{boe98}. 
If protons are accelerated efficiently in GRB blast waves, the spectrum in
the $\sim 10$ -- $100$~GeV range  is expected to be dominated by a hard
power-law with $\nu F_{\nu}  \propto \nu^{+0.5}$ due to proton synchrotron
radiation.  Since protons are expected to cool inefficiently, while
electrons  suffer strong radiative cooling, the temporal decay of the 
proton synchrotron radiation is slower than for the synchrotron component 
component. Let $g$ be the index parametrizing the deceleration 
and thereby the radiative regime of the blast wave, then both 
components decay as $F_{\nu} (t) \propto t^{-\chi}$, but with 
different temporal indices.  For synchrotron radiation 
from cooling electrons, $\chi_{\rm sy} = (4 g - 2)/(1 + 2 g)$, 
while for uncooled syncrotron radiation from UHECR, $\chi_{\rm p,sy} 
= (4 g - 3) / (1 + 2 g)$. The temporal decay of the SSC radiation 
is more complicated, because scattering in the Klein-Nishina 
is important. Comparison of decay observations over a large 
range of photon energies with model calculations are necessary 
to accurately discriminate between the various processes.
The SSL radiation can be  distinguished from the SSC and the hadronic
$\gamma$-ray  emission by its rapid variability and by the flares it
produces in the GeV range.  Observations of flares of GeV - TeV  radiation
would support the hypothesis that GRBs occur within stellar associations
and star-forming regions.

\end{document}